\documentclass[aps,prl,twocolumn,epsfig,amsmath,amssymb]{revtex4}
\usepackage[english]{babel} \usepackage{latexsym}
\usepackage{graphicx} \usepackage{subfigure} \usepackage{epsfig}
\usepackage{psfrag}

\begin{document}

\title{Resonant spin-changing collisions in spinor Fermi gases}
\author{N. Bornemann, P. Hyllus and L. Santos} 
\affiliation
{
\mbox{Institut f\"ur Theoretische Physik , Leibniz Universit\"at
Hannover, Appelstr. 2, D-30167, Hannover, Germany}
}

\begin{abstract}  
%
%

Spin-changing collisions in trapped Fermi 
gases may acquire a resonant character due to the 
compensation of quadratic Zeeman effect and trap energy. These resonances 
are absent in spinor condensates and pseudo-spin-1/2 Fermi gases, being 
a characteristic feature of high-spin Fermi gases that allows spinor physics 
at large magnetic fields. We analyze these resonances in detail 
for the case of lattice spinor fermions, showing that they permit  
to selectively target a spin-changing channel while suppressing all others. 
These resonances allow for the controlled creation of non-trivial
quantum superpositions of many-particle states with entangled spin 
and trap degrees of freedom, which remarkably are magnetic-field 
insensitive. Finally, we show that the intersite tunneling may lead
to a quantum phase transition described by an effective quantum Ising model.

\end{abstract}  
 \maketitle


Cold spinor gases have recently attracted 
a large interest.  A spinor gas is formed by atoms in two or more internal states 
simultaneously confined by optical traps \cite{Stenger98}.
Particularly intense efforts have been devoted to  
spinor Bose-Einstein condensates (BECs), which present a rich variety of 
ground state phases with different topologies \cite{Ho,spin1,spin2,spin3,classification}, 
such as ferromagnetic, polar, uni- and bi-axial nematic, and more.  
Spinor Fermi gases are also attracting currently a growing attention.
Collective modes in high-spin fermions have been shown to 
present rich features \cite{Yip-Ho}. Spin-$1$ Fermions allow for 
color superfluidity and baryon formation, linking 
spinor fermion physics to QCD \cite{Rapp2007}. Spin-$3/2$ fermions also present fascinating 
properties such as quintet Cooper pairing and Alice strings \cite{Wureview}.

The dynamics of spinor BECs has been also actively
investigated, in particular the coherent oscillations between  
spinor components \cite{Dynamics}. In addition, spinor gases 
constitute a novel tool for the analysis of out-of-equilibrium systems, 
and the generation of topological defects after a rapid quenching 
across a phase transition \cite{Kurn-Zurek}. On the contrary the spinor dynamics 
of high-spin fermions has been, to the best of our knowledge, not studied in
detail.

An external magnetic field typically constitutes a major
handicap for the spinor dynamics. 
Although, due to spin conservation during a collision, the linear Zeeman
effect (LZE) is irrelevant for the spinor dynamics, 
the quadratic Zeeman effect (QZE) may become (even for relatively low fields) sufficient to suppress spin change collisions, and hence 
any spinor dynamics \cite{Dynamics}. However, under appropriate conditions an external magnetic 
field may even stimulate spin dynamics in spinor BECs, due to the compensation of QZE and 
mean-field shifts \cite{Kronjager2006}. 

In the following, we show that spinor fermions allow for 
controllable magnetically-tuned resonances in the spin-changing dynamics,
 by compensating QZE and trap energy. These 
resonances are typically absent in spinor BECs, either because the mean-field energy 
greatly overwhelms the trap energy in the Thomas-Fermi regime, or because 
in the weakly-interacting regime the trap levels are irrelevant 
since the bosons mainly occupy the ground state of the trap. In lattice arrangements, as those discussed below, 
low temperature bosons occupy the lowest energy band, and hence the resonances
play no role. They are also absent in pseudo-spin-$1/2$
fermions, since due to spin conservation the QZE energy is conserved. 
On the contrary, these resonances constitute a
characteristic feature of high-spin trapped Fermi gases.

In this Letter we illustrate the role of these resonances for the case 
of fermions with spin $f=(2s+1)/2$ ($s\ge 1$) 
in an optical lattice, where each site acts 
as an independent anharmonic trap, with up to three relevant
levels (filling factor per spin component $\bar n<3$). 
However, we stress that the compensation 
between QZE and trap energy should also play an important role 
in more general optical traps, leading to trap-dependent resonant spinor dynamics, 
which will be the subject of a forthcoming work. Here, we study
these resonances in detail for the lattice case, including interaction-induced shifts and resonance
splittings, analyzing how for $f\ge 5/2$ these resonances may be employed to
select particular spin-changing collisions while suppressing all
others. Additionally, we show that an adiabatic sweep through the resonances
allows for the controlled creation of quantum superpositions of states with
entangled spin and trap degrees of freedom. These states are magnetic-field
insensitive, and hence, contrary to the usual case, these systems allow for
the study of spinor physics at large magnetic fields. Finally, we discuss the
effect of the inter-site tunneling, showing that under proper conditions it
may be described by a quantum Ising Hamiltonian, hence leading to a quantum
phase transition between different multiparticle states as a function of the
tunneling rate.

The independent sites are described by the Hamiltonian $\hat H=\hat H_0+\hat H_B+\hat H_I$. 
The trap energy is given by
\begin{equation}
  H_0=\sum_{n=0}^2 E(n) \sum_{m=-f}^f 
\hat a_{n,m}^\dag \hat a_{n,m},
\label{H0}
\end{equation}
where $E(n)=\hbar\omega (n+\beta n^2)$, and $\hat a_{n,m}$ is the annihilation operator of fermions 
of spin $m$ in the trap level $n$. The effective trap 
frequency ($\omega$) and anharmonicity ($\beta$) 
are obtained from the calculated first three on-site energies. 
The state of the system is described by a Fock state of the form:
$|N_{0,-f},\dots,N_{2,f}\rangle\equiv
\prod_{n=0}^2 \prod_{m=-f}^f (a_{n,m}^\dag)^{N_{n,m}}$, where 
we keep the order (from left to right) from $(n=0,m=-f)$ to $(n=2,m=f)$
\cite{footnote0}.

The effects of a magnetic field $B$ are given by
\begin{equation}
\hat H_B=\sum_{n,m} (-pm+qm^2)\hat a_{n,m}^\dag \hat a_{n,m},
\end{equation}
where the LZE is given by 
$p=g \mu_B B$, with $g$ the Land\'e factor and 
$\mu_B$ the Bohr magneton, and the QZE by 
$q=\mu_B^2B^2/4\hbar\omega_{hs}$, where $\omega_{hs}$ is 
the hyperfine splitting. 

Finally, the interatomic interactions (which we consider as being 
dominantly of short-range character) are provided by a Hamiltonian 
of the form \cite{Ho}
\begin{equation}
\hat H_I=\frac{1}{2}\sum_{\vec n,\vec m}C_{\vec n}U_{\vec m}
\hat a_{n_4,m_4}^\dag \hat a_{n_3,m_3}^\dag \hat a_{n_2,m_2} \hat a_{n_1,m_1},
\end{equation}
where $\vec n=\{ n_1,n_2,n_3,n_4 \}$, $\vec m=\{ m_1,m_2,m_3,m_4 \}$, 
$
C_{\vec n}=\int d^3 r \phi_{n_4}(\vec r)^*\phi_{n_3}(\vec r)^*
\phi_{n_2}(\vec r)\phi_{n_1}(\vec r),
$
and
$
U_{\vec m}=\sum_{F,M} g_F
\langle f,m_1,m_2|FM\rangle \langle FM|f,m_3,m_4\rangle .
$
In these expressions, $g_F=4\pi\hbar^2 a_F/m$, with $a_F$ the scattering length 
for the collisional channel with total spin $F$ (which 
due to symmetry must be an even number), $M=-F,\dots,F$, 
$\langle f,m_1,m_2|FM\rangle$ are the Clebsch-Gordan coefficients, 
and $\phi_n(\vec r)$ the $n$-th trap eigenfunction. The dynamics
is calculated by integrating the corresponding many-body Schr\"odinger equation, using
Runge-Kutta techniques.

\begin{figure}[ht] 
\begin{center}
\vspace*{0.1cm} 
\includegraphics[width=8.0cm]{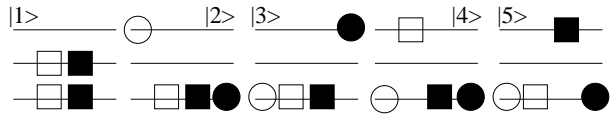}
\end{center} 
\caption{
States involved in the resonant dynamics at $\alpha\simeq -1/2$ for $f=3/2$. 
The lines indicate from bottom to top the 
three trap levels. The spin states are indicated by 
filled and hollow squares ($m=\pm 1/2$) and circles ($m=\pm 3/2$).}
\label{fig:1}
\vspace*{-0.2cm}  
\end{figure}

Although the resonances discussed below should be observable for general initial conditions, 
the analysis of the effects implied is simplified 
by considering an initial mixture of $m=\pm 1/2$ (at the end of this Letter we
discuss how this mixture may be achieved in on-going experiments). In addition, this initial state 
may be employed, as shown below, to controllably create complex superpositions of   
spin-level entangled states. For a filling factor (per component) $\bar n=n_0+\delta n$, 
with $n_0=0,1,\dots$, all levels up to $n_0-1$ are filled in all
sites, whereas the level $n_0$  is occupied with 
a probability $\delta n$. This imprecise filling becomes eventually important, and 
is taken into account in our simulations. For $\bar n=2$, state $|1\rangle$ in
Fig.~\ref{fig:1} is the initial state for the dynamics.

A collision leads to a QZE shift $\Delta E_{QZE}=q(m_3^2+m_4^2-m_1^2-m_2^2)$. 
As discussed above, this shift prevents spin-change collisions, 
as long as $|\Delta E_{QZE}|$ is larger than the typical 
interaction energy of spin-change collisions.
Level-changing collisions are characterized by an 
energy shift $\Delta E_{TRAP}=E(n_4)+E(n_3)-E(n_2)-E(n_1)$. 
An interesting situation occurs when spin-changing collisions 
are at the same time level-changing. In that case, $\Delta E_{QZE}$ and 
$\Delta E_{TRAP}$ may compensate each other, leading to resonances in the spin-changing collisions. 
Spin conservation precludes these resonances in usual spin-$1/2$ Fermi systems, which thus do not show 
level-changing collisions if $|\Delta E_{TRAP}|$ is larger than the 
spin-changing interaction energy. In the following, and for simplicity of our discussion, 
we concentrate in situations where the harmonic trap energy can be assumed as conserved 
($n_1+n_2=n_3+n_4$), and only the anharmonic change 
$\Delta E_{ANH}=\beta\hbar\omega (n_3^2+n_4^2-n_1^2-n_2^2)$ is relevant. 
In general, however, a sufficiently large B may lead to 
resonances which involve a violation of the previous condition, leading to a 
wealth of resonances and complex multiparticle superpositions, similar to those discussed below. 
%
%

\begin{figure}[ht] 
\begin{center}
\includegraphics[width=5.5cm,angle=0]{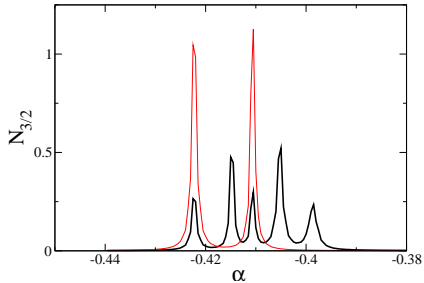}
\end{center} 
\vspace*{-0.2cm} 
\caption{(color online) 
$f=3/2$. Time-averaged population ($N_{3/2}$) in $m=\pm 3/2$
as a function of $\alpha$, for $\tilde g_{0,2}=0.08, 0.10$~\cite{footnote1}, 
and $\bar n=2.0$ (thin, red), and $2.5$ (thick). 
}
\label{fig:2}
\vspace*{-0.2cm}  
\end{figure}

The resonances are altered by interparticle interactions. 
We illustrate this by considering the initial $f=3/2$ state $|1\rangle$ 
 (see Fig.\ref{fig:1}).
Note that multi-particle
states $|2\rangle$ to $|5\rangle$ posses equal QZE plus anharmonic energy, 
$E_{QZE}+E_{ANH}=5q+4\beta\hbar\omega$ (an important point as
discussed below), whereas for $|1\rangle$
$E_{QZE}+E_{ANH}=q+2\beta\hbar\omega$. Hence a resonance would be expected 
at $\alpha\equiv \frac{q}{\beta\hbar\omega} =-1/2$. However, a detailed analysis shows that close to resonance the states 
indicated in Fig.~\ref{fig:1} are isolated from the rest of all other possible (seven) states
that can be reached via $\hat H_I$ from $|1\rangle$, 
leading to an effective $5$-state problem. Interestingly, one may  
show that $|D_1\rangle\equiv (|3\rangle+|2\rangle)/\sqrt{2}$, and
$|D_2\rangle\equiv (|5\rangle+|4\rangle)/\sqrt{2}$ decouple from $|1\rangle$, 
playing a similar role as dark-states 
in quantum optics. The dynamics is then fully dominated by the coupling of 
$|1\rangle$ and two ``bright'' states 
$|\pm\rangle=\frac{1}{2}[(|3\rangle-|2\rangle)\pm (|5\rangle-|4\rangle)]$. 
As a consequence, the resonance splits into two peaks, corresponding to the 
energy shifts between $|1\rangle$ and $|\pm\rangle$:
\begin{equation}
\alpha_R=-1/2+\left [
  g_0\xi_0+g_2\left (\xi_1\pm\delta\xi\right ) \right ]/4\beta\hbar\omega
\end{equation}
where $\xi_0=C_{1100}-C_{2200}+C_{1111}/2$, $\xi_1=-2C_{0000}-3C_{2200}+C_{1100}+C_{1111}/2$, and 
$\delta\xi=2C_{2200}$. 
Note that, due to the EPR-like nature of the spin-change collisions, 
the resonances are accompanied by transitions into   
quantum superpositions of multi-particle states with entangled spin and trap level
degrees of freedom, which we explore below.
Split resonances for $1<\bar n < 2$ (Fig.~\ref{fig:2}) are hence  
a direct consequence of the coherent formation of the above mentioned many-body entangled states.
The split resonance peak may be employed to selectively detect $|\pm\rangle$
(which may become important to probe the linear superpositions discussed
below), since initial $|+\rangle$ or $|-\rangle$ states lead to 
different shifts of the resonance into $|1\rangle$, observable 
by monitoring the $m=1/2$ population. The picture gets 
more complicated if $2<\bar n < 3$ (Fig.~\ref{fig:2}), 
since the resonances are shifted depending 
whether there is none, one or two particles in the third trap level. Multiple 
resonance peaks appear (or a resonance broadening if the individual peaks are not resolved), 
but in this case due to the imprecise filling, as well as to the above
mentioned splitting. Since the spinor dynamics depends
on $\bar n$, one observes plateaux in $d^2 N_{3/2}/d\bar n^2$ (where $N_{3/2}$ is the 
sum of the time-averaged populations in $m=\pm 3/2$), with jumps 
when $\bar n$ becomes an integer (resembling 
the de Haas-van Alphen susceptibility plateaux).


\begin{figure}[ht] 
\begin{center}
\includegraphics[width=6.0cm,angle=0]{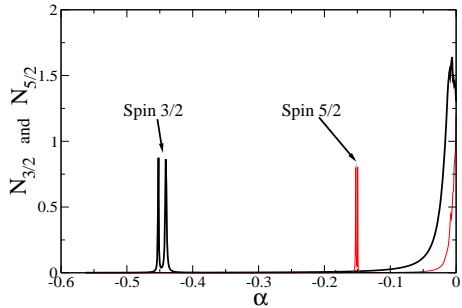}
\end{center} 
\vspace*{-0.2cm} 
\caption{(color online) $f=5/2$. Time averaged population $N_{3/2}$ 
in $m=\pm 3/2$ (thick) and 
$N_{5/2}$ in $m=\pm 5/2$ (thin, red) as a function of $\alpha$, for $\bar
n=2.0$, and $\tilde g_{0,2,4}=0.08, 0.10,0.05$~\cite{footnote1}.
}
\label{fig:3}
\vspace*{-0.2cm}  
\end{figure}

Whereas for $f=3/2$ only one spin-changing collisional channel is possible, 
$f\ge 5/2$ allows for more channels.
Fig.~\ref{fig:3} shows the case of spin-$5/2$. Note that with an initial state
$|1\rangle$ only two spin-changing collisions are possible:
from $(n_{1,2}=1;m_{1,2}=\pm 1/2)$ either into $(n_{3,4}=0,2;m_{3,4}=\pm 3/2)$ or 
$(n_{3,4}=0,2;m_{3,4}=\pm 5/2)$, with respective resonances at 
$\alpha_R\simeq -1/2$ and $\alpha_R\simeq -1/6$. 
Interestingly, if the two resonances are well resolved, as it is the case in Fig.~\ref{fig:3}, 
it may be possible to select a particular spin-changing channel, preventing 
all others, providing a novel tool for the manipulation of spinor gases. 
As for the $f=3/2$, the resonances couple the initial states with non
trivial superpositions of spin-level entangled states.


As previously mentioned, quantum superpositions of spin-level entangled
states play an important role in the dynamics at the spin-changing
resonances. These states may be created in a controllable way by sweeping 
adiabatically the external magnetic field across the resonances. As an
example, we consider the spin-$3/2$ case. If $\beta<0$, the state $|1\rangle$
is the ground-state for $|\alpha|>|\alpha_R|$. In that case, 
if $g_2>0$ ($g_2<0$) an adiabatic sweep towards $|\alpha|<|\alpha_R|$ 
creates $|+\rangle$ ($|-\rangle$), with a fidelity larger than
$95\%$ for sweeps with $d\alpha/dt=2.5\times 10^{-4} \frac{\omega}{2\pi}$. If
the gas is prepared initially at $\alpha=\alpha_R-0.05$, this linear ramp
towards $\alpha=\alpha_R+0.05$ demands a change in $4000$ trap periods, which
for typical values of the effective on-site $\omega$ (few $kHz$) represents sweep times of less than
$0.5$s. Note that for $\beta<0$, the created states are the ground-states for
$|\alpha|<|\alpha_R|$, except for $g_2<0,g_0>-3g_2$ and $g_2>0,g_0>g_2$, where $D_{1,2}$ are 
the (degenerate) ground states. However, as mentioned above, these states are dark states 
and hence they are not coupled to $|1\rangle$ during the sweep. 
Note that remarkably, all states belonging to the superpositions $|\pm\rangle$
possess the same QZE. This leads to two important consequences. On one hand, 
the sweep across the resonance is produced at rather large $B$ 
($\sim$ 1 G, see below), 
but the result of the sweep crucially depends on the spinor physics (sign of
$g_2$). Hence spinor physics becomes relevant even 
for large $B$. Second, once the states are produced an abrupt 
increase of $B$ leaves the system unchanged, and hence the 
superpositions created may be robust at very large $B$, even opening the
possibility of employing Feshbach resonances to variate $g_2$ or
$g_0$, without polarizing the system due to the QZE. Similar states may be
created by adiabatic sweeps for $f>3/2$. E.g., for $f=5/2$ a sweep would create a 
state 
$\cos(\phi)(|2\rangle -|3\rangle)/\sqrt{2}+\sin(\phi)(|4\rangle
-|5\rangle)/\sqrt{2}$, where
$\tan\phi=(\epsilon-\sqrt{\epsilon^2+\Omega^2})/\Omega$, 
$\epsilon=15(2g_2-g_4)(C_{0220}-C_{0000})/42$, and 
$\Omega=2(9g_4-2g_2-7g_0)C_{1102}/7$.


The states discussed above are created in isolated lattice sites. 
However, if the
lattice is relaxed, the tunneling for the third band may become
relevant, especially if the splitting $4g_2C_{0220}$ between the states $|\pm\rangle$ 
becomes very small. In that case, different types of lattice Hamiltonians may 
be created. Let us consider the particular case of spin-$3/2$, with $|g_2|\ll |g_0|$. 
In that case, the states $|\pm\rangle$ form a quasidegenerate 
pseudo-spin-$1/2$ 
doublet. 
If the tunneling $t$ for the third band satisfies $t\ll |g_0|$, then 
second order processes lead to an effective spin Hamiltonian. 
Note that the condition $|g_2|\ll |g_0|$ is necessary to avoid mixing other 
possible states. 
Employing 
$|+\rangle\langle +|=1/2+\hat S^z$,
$|-\rangle\langle -|=1/2-\hat S^z$,
$|+\rangle\langle -|=\hat S^x+i\hat S^y$,
$|-\rangle\langle +|=\hat S^x-i\hat S^y$,
the effective lattice Hamiltonian may be reduced to a quantum Ising
Hamiltonian \cite{Sachdev}
\begin{equation}
\hat H_{eff}=-J_{eff}\sum_i\hat S_i^z-W_{eff}\sum_{\langle i,j\rangle}\hat
S_i^x \hat S_j^x
\end{equation}
where $J_{eff}=4g_2 C_{0220}$, and $W_{eff}=\frac{4t^2}{3g_0C_{0220}}
\frac{2C_{2222}-3C_{0220}}{C_{2222}-3C_{0220}}$. Typically
$W_{eff}>0$ ($<0$) if $g_0>0$ ($<0$). Hence there is a critical tunneling, 
$t_c^2=3g_2g_0 C_{0220}^2 \left (
\frac{C_{2222}-3C_{0220}}{2C_{2222}-3C_{0220}} \right )$, such that for 
$t<t_c$ the single-site physics dominates, and $|+\rangle$ ($|-\rangle$) 
is the ground-state for $g_2>0$ ($g_2<0$). For $t>t_c$ the tunneling
dominates, and for $g_0>0$ the system enters into a ferromagnetic state 
with all sites in either $|u\rangle\equiv (|3\rangle-|2\rangle)/\sqrt{2}$ or 
$|d\rangle\equiv (|5\rangle -|4\rangle)/\sqrt{2}$. If $g_0<0$ the coupling 
is antiferromagnetic, and hence a 1D system enters into a staggered
configuration with alternated $|u\rangle$ and $|d\rangle$ states. The latter
case opens the possibility for the analysis of frustration in other lattice geometries.


Finally, we comment on possible experimental realizations.
We stress that the resonances should play an important role in
high-spin Fermi gases in general dipole traps. For the case 
discussed above, the resonances should be clearly
observable in standard lattices, $V_0 \sin^2 \kappa x$. However, these lattices 
may be inappropriate for the detailed study of the rich phenomena discussed, since 
very large $V_0> 50 E_{rec}$ (where $E_{rec}=\hbar^2\kappa^2/2m$)
is needed to isolate the three lowest levels at each site from other sites, and 
it may be difficult to load only the first three 
bands of an initially weak lattice, due to 
the vanishing gap $\Delta E_{34}$ between the third and fourth band. 
A better scenario is provided by trimerized lattices (three wells per
elementary cell) \cite{Santos04}. In our calculations we have 
considered for simplicity a 1D trimerized lattice 
$V(x)=V_0 (\sin^2(\kappa x/3)+\frac{1}{2}\cos^2(2\kappa x/3)+\frac{3}{4}\sin^2(\kappa x)$, 
assuming a strong confinement in the $yz$-plane provided e.g. by an additional
2D lattice. $V(x)$ allows for a relatively large $\Delta E_{34}$ even 
for relatively weak lattices. E.g. for $V_0=0.5 E_{rec}$, 
$\Delta E_{34}\simeq 0.22 E_{rec}$ (for, e.g., $^{40}K$, typically $E_{rec}/k_B\simeq 2\mu$K), 
whereas the tunneling for the lowest bands is still sufficiently large
(allowing for thermalization in the weak lattice). In this way, it may be possible 
to prepare controllably a Fermi gas in the first three bands, 
if the temperature $T<\Delta E_{34}$ and $\bar n<3$. 
The considered initial $m=\pm 1/2$ mixture may be created from a polarized Fermi gas 
in a maximally stretched state, by first employing an adiabatic 
radio-frequency sweep to transfer into $m=1/2$, and then a $\pi/2$ radio-frequency 
pulse to establish a coherent $m=\pm 1/2$ mixture. 
A sufficiently large B (away from the discussed resonances) 
isolates this mixture against spin-change collisions
due to QZE. In the presence of polarized bosons (e.g. in  
KRb mixtures), this mixture would thermalize in the weak lattice 
leading to an incoherent $m=\pm 1/2$ mixture, which  after a lattice ramp-up 
would lead to the isolated sites discussed in this Letter. In a last stage 
the bosons should be eliminated, and $B$ brought to the resonant values. 
For $^{40}K$, $q/\hbar\simeq 70 \left (\frac{B}{{\rm
    G}} \right )^2$ Hz, hence for $2\pi/\kappa=800$nm 
and  $V_0=3.3 E_{rec}$ ($\omega/2\pi\simeq 9.25$kHz, $\beta=-0.3$),
$\alpha=-1/2$ occurs for $B\simeq 4.45$G. 

Summarizing, spin-changing collisions in Fermi gases become resonant if 
the QZE compensates the trap energy. These resonances are absent in BECs and spin-$1/2$ Fermi gases, being 
a characteristic feature of high-spin fermions.
We have shown for the case of lattice fermions that the 
resonances are shifted and split by interactions, and may permit 
to target a single spin-changing channel while avoiding
the rest. The resonances allow for the controlled creation of quantum 
superpositions of states with entangled spin and trap-level degrees of
freedom. These superpositions are 
magnetically isolated, and hence these systems allow for the observation
of spinor physics at large magnetic fields, in principle even at Feshbach resonances. 
Finally, tunneling may compete with the on-site spinor
physics to lead to a quantum phase transition described by a quantum Ising
model. The rich phenomenology deriving from these resonances is by no
means exhausted by the discussion above. Interesting physics
is expected in the spinor dynamics of high-spin fermions in dipole traps. In
addition, different resonances and spin mixtures in lattice fermions may allow
for a wealth of  
quantum superpositions of spin-level entangled states, and 
various types of effective Hamiltonians. These perspectives will be the
focus of forthcoming works.

\acknowledgements
We thank C. Klempt, Th. Henninger, and J. Arlt for useful discussions. 
This work was supported by the DFG (SFB-TR21, SFB407, SPP1116).

\end{document}